\documentstyle[aps,multicol]{revtex}
\input{epsf}
\def\arQ{Q}
\def\spv{{\bf n}}
\def\Spv{{\bf N}}
\def\pov{{\bf R}}
\def\sov{{\bf g}}
\def\phT{\phi_{\rm T}}
\def\phQ{\phi_{0}}
\def\pcd{q_{\rm P}}
\begin{document} 
\draft 
\title{Magnetotransport in manganites and 
the role of quantal phases I: Theory}
\author{Y.~Lyanda-Geller, P.~M.~Goldbart, 
        S.~H.~Chun and M.~B.~Salamon} 
\address{Department of Physics and Materials Research Laboratory, \\  
University of Illinois at Urbana-Champaign, Urbana, Illinois 61801}
  \date{April 22, 1999}
\maketitle
\begin{abstract} 
A microscopic picture of charge transport in manganites is developed,
with particular attention being paid to the neighborhood of the 
ferromagnet-to-paramagnet phase transition. The basic transport 
mechanism invoked is inelastically-assisted carrier hopping between 
states localized by magnetic disorder. In the context of the 
anomalous Hall effect, central roles are played by the 
Pancharatnam and spin-orbit quantal phases.
\end{abstract}
\begin{multicols}{2}
\narrowtext
\noindent
{\sl Introduction\/}: 
The double-exchange interaction (DEI) has long been understood to play a
major role in the ferromagnet to paramagnet transition (FPT) in the
manganite systems La$_{1-x}$A$_x$MnO$_3$ (where A stands for Ca, Sr or
Pb), the transition being accompanied by a metal-insulator transition
(MIT). In this DEI picture, proposed by Zener and elaborated by Anderson
and Hasegawa~\cite{Anderson}, intra-atomic Hund's Rule coupling leads to
a modulation of the amplitude for the hopping of outer-shell carriers
between neighboring Mn ions.  It is now recognized, however,
that the physics of the DEI is insufficient to
fully explain the observed phenomenon of colossal magnetoresistance
(CMR) (i.e., the strong magnetic-field induced
suppression of the resistivity, and the shift to higher temperatures of the
peak in its temperature-dependence~\cite{Ramirez}). 
Moreover, interest in CMR has led
to a re-examination of the nature of the FPT and the MIT in manganites
and related compounds.  In these contexts, Millis, Shraiman and
co-workers~\cite{Millis1,Millis2} have proposed that the DEI is
accompanied by a large Jahn-Teller lattice distortion that would cause
the polaronic collapse of any conduction band. Varma~\cite{localization1}
and Sheng et al.~\cite{localization2} have argued, in contrast, that the
MIT in manganites is an Anderson localization transition, resulting from
magnetic and nonmagnetic disorder.

The purpose of the present Letter is to address charge 
transport in manganites in the vicinity of the FPT and MIT 
from the vantage point afforded by the Hall effect.  In a  
companion Letter~\cite{REF:Salamon}, we present and analyze experimental 
data on the Hall effect and CMR in La$_{2/3}$(Ca,Pb)$_{1/3}$MnO$_3$. 
We shall argue that, near the FPT, owing to charge-carrier localization, 
transport is via hopping between localized states. 

The central part of our analysis is the discussion of the microscopic
mechanism of the Hall effect (HE) in manganites. In ferromagnetic metals 
HE's include
an {\it ordinary\/} Hall effect (an OHE, which arises 
from the Lorentz force acting on the current carriers),  as well as an 
{\it anomalous\/} Hall effect (AHE), i.e., a Hall current proportional to
the average magnetization and independent of demagnetization effects.  
For metallic states, microscopic mechanisms
yielding the AHE have been discussed, e.g., in Ref.~\cite{Luttinger},
the essential ingredient being the spin-orbit interaction (SOI), 
which leads to an AH current in the presence of magnetization
(of any origin)~\cite{so}.  If charge
transport near the FPT and MIT in manganites does indeed occur via
hopping, then we are led to the general issue of the microscopic
mechanism of the AHE in hopping conductors.  This AHE cannot be captured
by a picture based solely on the Anderson-Hasegawa
analysis~\cite{Anderson} of the DEI within a {\it pair\/} of Mn ions.  
Such a picture includes only the modulation of the {\it magnitude\/} 
of the hopping between the pair determined by the
relative alignment of the core spins on the ions [via a factor
$\cos(\theta/2)$, where $\theta$ is the angle between (semiclassical)
directions of the core spins].  This insufficiency of a pair-based 
picture is an analog of Holstein's observation~\cite{REF:Holstein1} 
that to capture the OHE in hopping conductors requires the analysis of 
{\it at least triads\/} of atoms, and of the attendant Aharonov-Bohm 
(AB) fluxes through the polygons whose vertices are the atomic sites.  
Therefore, we shall examine a mechanism for the AHE involving hopping 
within {\it triads\/} of sites, in which fundamental roles 
are played by {\it two quantal phases\/}: (i)~the SOI phase, acquired 
by electrons propagating in the presence of SOI; and (ii)~the (quantal) 
Pancharatnam phase (an electronic analog of the (optical) Pancharatnam 
phase accrued by classical light under a sequence of polarization 
changes~\cite{REF:Pancha,REF:BonPan}).  
In this electronic analog, outer-shell carriers, hopping from ion to ion,
acquire a phase~\cite{footnote} determined by the solid angle subtended
by the spherical polygon whose vertices are the orientations of the
core-spins of the ions visited.

Recently, Kim et al.~\cite{REF:Kim} revisited the theory of the AHE, in
the context of a model that includes DE, SOI, and gauge fluxes arising 
from interactions.  In work done in parallel with the
present work, Ye et al.~\cite{REF:Ye1999}, focusing on the metallic regime, 
address the
relationship between the AHE, Berry phases~\cite{footnote} and the SOI, 
and like the present work, incorporate the effect of topological 
spin excitations.

\noindent
{\sl Localization of carrier states in manganites\/}: 
Several general ideas support the notion that the carrier states are 
localized at temperatures near to the (zero magnetic field) FPT, as 
well as at higher temperatures.  Approaching the FPT from the 
ferromagnetic side, there is a net magnetization of the core spins, 
but strong thermal fluctuations render typical instantaneous 
configurations of the spins rather inhomogeneous.  Among these 
fluctuations there are \lq\lq hedgehog\rq\rq\ excitations which, 
owing to their topological stability, are long-lived, and become 
more numerous as the FPT is approached~\cite{Lau}.  Due to the 
resulting inhomogeneity, the carrier-transfer matrix elements are 
reduced~\cite{REF:shrink}.  In the quasi-static approach, the (fast) 
carrier motion takes place through a slowly (time-)varying background
core-spin configuration. In generic instantaneous random backgrounds, the
carriers are expected to be localized.  Support for this notion comes
from the close similarity between manganites and a system of randomly
located identical impurities (i.e., off-diagonal disorder) for which 
localization has been established by Lifshitz~\cite{REF:Lifshitz}.
Although spin-induced randomness in manganites [arising from the random
$\cos(\theta/2)$ factors] is weaker than the randomness considered 
in~\cite{REF:Lifshitz}, 
the two systems are expected to exhibit similar localization 
behavior.  Furthermore, the condition for localization (viz., that the
characteristic spatial scale of the outer-shell wavefunctions be
much smaller than distance between sites) is well obeyed in manganites.
Therefore, provided that there is appreciable randomness in the 
core-spins orientations, the transport properties should be determined by 
the short-distance physics of clusters of ions and magnetic correlations 
between such clusters.  Moreover, nonmagnetic disorder and possible states 
bound to the A-ions are capable of amplifying the trend towards
localization~\cite{localization1,localization2}.

Thus the following picture of transport in manganites emerges.
(i)~In the paramagnetic insulating state, the percolative motion of 
strongly localized carriers is suppressed by magnetic randomness.  
(ii)~With decreasing temperature, the carrier hopping (which is assisted 
by phonons) becomes less frequent, so that the resistivity grows, and 
(iii)~reaches a maximum when the core spins become sufficiently 
correlated that a tenuous but infinite conducting network emerges. 
(iv)~With further reduction in temperature, the resistivity 
decreases abruptly, in line with the traditional percolation 
picture~\cite{REF:AHL}, as more and more hopping paths 
become available to carriers, owing to further alignment of core spins.  
This abrupt decrease terminates when the newly available hopping paths 
are effectively shunted by the existing network. (v)~Further decrease in 
temperature leads to further core-spin alignment and, ultimately, to 
the metallic state. 

  \begin{figure}[hbt]
  \epsfxsize=4.5truecm
   \centerline{\epsfbox{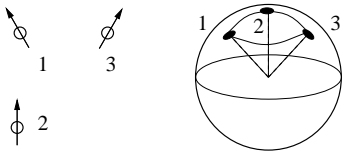}} 
  \vskip+0.20truecm
  \caption{(a)~Triad of Mn ions. 
          (b)~Spherical triangle with core-spin 
              orientations as vertices.}
  \label{FIG:one}
  \end{figure}
\noindent
{\sl Anomalous Hall effect and the Pancharatnam phase\/}: 
In order to discuss the AHE in conditions of charge-carrier localization, 
we begin by considering a triad of magnetic sites formed by neighboring 
Mn ions, as shown in Fig.~\ref{FIG:one}.  Within such triads, there is 
an elementary AHE, which arises from interference between hopping processes 
connecting two sites: e.g., between the direct process (having complex 
amplitude ${\cal A}$) and the indirect process of hopping via the third site 
(with amplitude ${\cal A}^{\prime}$).  Ignoring any AB flux (as we shall 
not be concerned with the OHE), we observe that any phase difference 
between the amplitudes ${\cal A}$ and ${\cal A}^{\prime}$ stems from 
spin quantal phases and transfer-assisting mechanisms (e.g., 
electron-phonon processes); we call the latter {\it transfer phases\/} 
$\phT$. 

To understand the nature of the spin quantal phases we examine the 
single-particle quantum mechanics of a carrier hole added to a triad 
of ${\rm Mn}^{3+}$ ions.  We regard the spin-3/2 core spins of the Mn 
ions as large enough to be treated classically, so that one can assign 
a definite direction to each.  Thus, a generic configuration is 
characterized by the unit vectors 
$\big\{\spv_{1},\spv_{2},\spv_{3}\big\}$ located at the triad of sites 
$\big\{\pov_{1},\pov_{2},\pov_{3}\big\}$ (see Fig.~\ref{FIG:one}).  
Due to Hund's Rules there is, at each site, a single state available 
to the added hole, its spin opposing the core-spin direction.  We treat 
the remaining spin (and orbital) states as simply being inaccessible.
Postponing to below the effects of SOI, we assume that the transfer
of holes (being effected by either the kinetic energy or the
electron-phonon interaction) has no effect on the spin of the carriers.
However, such transfer in the presence of the constraints set by the
core-spin orientations has a striking effect on the quantal dynamics of
the carriers: in the quantal amplitude for a hole to move
once around the triad, 
viz.~${\cal A}^{\prime\,\ast}\,{\cal A}
\propto{\rm Tr}\,P_3\,P_2\,P_1$ 
[where the operator 
$P_j\equiv(1+\mbox{\boldmath$\sigma$}\cdot{\bf n}_j)/2$ 
projects onto the spinor aligned with $\spv_{j}$], 
there arises a quantal phase,
\[
\Omega/2\!=\!\tan^{-1}
\big[
\spv_{1}\!\cdot\!(\spv_{2}\!\times\!\spv_{3})/\!
\left(1\!+\!\spv_{1}\!\cdot\!\spv_{2}\!+\! 
\spv_{2}\!\cdot\!\spv_{3}\!+\!\spv_{3}\!\cdot\!\spv_{1}\right)\!\big],
\]
which modulates the interference between direct and indirect hopping 
between sites of a triad.  $\Omega$ is the (oriented) solid angle of 
the geodesic triangle on the unit sphere having vertices at 
$\{\spv_1,\spv_2,\spv_3\}$.  It is the quantal analogue of the 
classical optical phase discovered in the context of polarized light 
by Pancharatnam~\cite{REF:Pancha,REF:BonPan}.  What Pancharatnam showed 
is that a cyclic change of the polarization state of light is accompanied 
by a phase shift (i.e., a phase anholonomy) determined by the geometry of 
the cycle, as represented on the Poincar\'e sphere of light polarizations, 
via the area $\Omega$ of the geodesic polygon whose vertices are these 
polarizations.

In the DE electronic analog, the transporting of a d-shell carrier to an
ion with a differently oriented core spin in a spin-independent process
amounts to a {\it connection\/}, which determines the phase of the spin
state in terms of the sequence of sites visited.  A hole returning to a
site returns to the same spin-state, except that its phase is augmented
by a quantal Pancharatnam phase, determined by the geometry of the
cycle, as represented on the sphere of core-spin orientations, via 
{\it half\/} the area of the geodesic polygon whose vertices are these
orientations.  In contrast to Berry's adiabatic phase~\cite{footnote},
the phenomenon described here is associated with {\it sudden\/} 
changes in the carrier-spin state, and need not be slow.

In the hopping regime, this Pancharatnam phase leads to an AHE in 
an elementary triad in much the same way that an AB flux leads to 
the OHE in Holstein's spinless model~\cite{REF:Holstein1}.  In 
Holstein's model, carrier hopping between sites of a triad occurs 
due to carrier-phonon interaction; the Hall current arises due to 
the interference of direct and indirect hopping.  The transfer 
phase is nontrivial ($\phT=\pi/2$) when this interference involves 
processes assisted by two phonons~\cite{REF:numphon}.  (For the 
longitudinal conductivity, a single phonon-assisted transfer is 
sufficient.)\thinspace\  In a uniform magnetic field ${\bf B}$, 
processes associated with a nontrivial $\phT$ lead to an OH 
conductivity~\cite{REF:Holstein1},
\begin{equation}
\sigma_{\rm OH}=
G\{\epsilon_{j}\}\,
\sin\phT\,
\sin\big({\bf B}\cdot{\bf\arQ}/\phQ\big),
\end{equation}
where $\phQ$ is the (electromagnetic) flux quantum, 
${\bf\arQ}$ is the (oriented, real space) area of the triad, and 
$\{\epsilon_{j}\}_{j=1}^{3}$ are the energies of the three single-particle 
eigenstates, which are invariant under reversal of the AB flux.  
The explicit expression for $G$ can be found in Ref.~\cite{REF:Holstein1}.  
($G$ also depends on the populations of these states, which 
themselves may depend on particle-particle correlations.)\thinspace\ 

We now turn from the OHE in a spinless triad to the elementary 
AHE in a triad of magnetic sites.  Like the OHE, this AHE results from 
two-phonon processes, but is due to the Pancharatnam phase instead of 
the AB phase.  (We do not yet included the effects of the SOI.)\thinspace\ 
{\it Mutatis mutandis\/}, we arrive at the AH conductivity, 
\begin{equation}
\sigma_{\rm AH}=
  G\{\varepsilon_{j}\}\,
\sin\phT\,
  \cos\frac{\theta_{13}}{2}\,
  \cos\frac{\theta_{32}}{2}\,
  \cos\frac{\theta_{21}}{2}\, 
  \sin\frac{\Omega}{2},
\label{EQ:AHcurrent}
\end{equation}
where $\cos\theta_{jk}\equiv\spv_{j}\cdot\spv_k$, 
$\cos(\theta_{jk}/2)$ are Anderson-Has{\-}egawa factors, and 
$\{\varepsilon_{j}\}$ are the energies of the three 
single-particle eigenstates consistent with Hund's Rules, 
these energies depending on 
  $\spv_{j}\cdot\spv_{k}$ 
and $\cos(\Omega/2)$.  Note that $G$ is invariant under 
Pancharatnam flux reversal $\Omega\to-\Omega$, and 
$\sigma_{\rm AH}$ is odd under it. 

We have shown that, for a triad with given set of core-spin orientations, 
an AHE arises from the quantal Pancharatnam flux.  However, there is a 
significant difference between this AHE and the OHE.  In the former 
(nonmagnetic) case, a uniform applied magnetic field leads to a net 
macroscopic OHE, even though contributions of triads may 
cancel one another~\cite{REF:Galperin}.  In the latter case (magnetic 
sites, Pancharatnam flux, and no SOI), even the presence of 
macroscopic magnetization of the core 
spins is insufficient to cause a {\it macroscopic\/} 
Hall current.  The reason for this is that in 
obtaining the macroscopic AH current from Eq.~(\ref{EQ:AHcurrent}) 
we must average over the configurations of the core spins.  In the 
absence of SOI, the distribution of these configurations, although 
favoring a preferred {\it direction\/} (i.e., the magnetization direction 
${\bf m}\equiv{\bf M}/M$), is invariant under a reflection of all 
core-spin vectors in any plane containing the magnetization.  This fact, 
coupled with the fact that $\{\varepsilon_{j}\}$ are also invariant 
under such reflections, guarantees that the macroscopic AH current will 
average to zero.  (We do, however, expect significant AH current 
{\it noise\/}, in the FPT regime, owing to the fluctuations of the 
Pontryagin charge~\cite{REF:IoLeMi} of the triads of core spins and, 
hence, elementary Pancharatnam fluxes.) 

In order to capture the AHE in materials such as manganites, 
we must consider not only the Pancharatnam phase but also 
some agent capable of lifting the reflection invariance of the 
energies $\{\varepsilon_{j}\}$ and the distribution of core-spin 
configurations, and hence of inducing sensitivity to the sign of the 
Pancharatnam flux.  Such an agent is provided by 
the SOI, 
$
H_{\rm so}=\alpha{\bf p}\cdot
\left(\mbox{\boldmath $\sigma$}\times
\mbox{\boldmath $\nabla$}U\right), 
$
where $U$ includes ionic and impurity potentials, $\alpha$ is the SOI 
constant, ${\bf p} $ is the electron momentum, and $\mbox{\boldmath 
$\sigma$}$ are the Pauli operators.  The SOI leads to an effective SU(2) 
gauge potential 
${\bf A}_{\rm so}=\alpha m (\mbox{\boldmath $\sigma$}
\times\mbox{\boldmath $\nabla$}U)$~\cite{Goldhaber}, providing an 
additional source of quantal phase.  For a given core-spin configuration, 
SOI favors one sense of carrier-circulation around the triad over the 
other, and thus favors one sign of Pancharatnam phase over the other. 
 
There are two resulting contributions to the AHE.  The first, 
$I^{(1)}_{\rm AH}$, arises from the SOI-generated dependence of 
$\{\varepsilon_{j}\}$ on the three vector-products 
$\Spv_{jk}\equiv\spv_{j}\times\spv_{k}$ 
which, together with the magnetization direction ${\bf m}$, yield a 
preferred value for the triad Pontryagin charge 
$\pcd$ [$\equiv\spv_{1}\cdot(\spv_{2}\times\spv_{3})$] and, hence, 
a preferred Pancharatnam flux.  To see the origin of this dependence 
on $\Spv_{jk}$, let us analyze corrections, due to the SOI, of hole 
eigenenergies.  If the on-site energies of the holes are nondegenerate, 
it is straightforward to determine that phase sensitivity first enters 
at third order (in the transfer matrix elements):
$\delta\varepsilon_j\!=\!\sum_{h,k(\ne j)} 
{\rm Tr}\,T_{jh}\,T_{hk}\,T_{kj}\big/
(\varepsilon_j\!-\!\varepsilon_h)(\varepsilon_j\!-\!\varepsilon_k)$, 
where $T_{jk}\equiv P_j\,V_{jk}\,P_k$ are the transfer amplitudes, 
$V_{jk}$ are the hopping matrix elements, and 
${\rm Tr}$ denotes a trace in spin space. (For degenerate $\varepsilon$'s 
one should obtain the splitting of the $\varepsilon$'s due to transfer in 
the absence of SOI, and then include SOI at the final step, arriving at 
the result to be given below.)\thinspace\  The hopping matrix elements 
are sensitive to the SOI quantal phase, and can be written in the form 
$V_{jk}= V_{jk}^{\rm orb}\,L_{jk}$, where 
$L_{jk}\equiv\big(1+i\mbox{\boldmath$\sigma$}\cdot{\bf g}_{jk})$, 
$V_{jk}^{\rm orb}$ is an orbital factor, and ${\bf g}_{jk}$ 
($\propto\alpha_{\rm so}$) is an appropriate vector that describes 
the average SOI for the transition $j\rightarrow k$~\cite{REF:magofG}.  
Then, e.g., the first-order (in $\alpha$) shifts in the 
$\varepsilon$'s are given by 
\begin{eqnarray}
&&
\delta\varepsilon_j\propto 
{\rm Tr}\,T_{13}\,T_{32}\,T_{21}=
4\,{\rm Re}\,{\rm Tr}\,
P_1\,L_{13}\,P_3\,L_{32}\,P_2\,L_{21}
\nonumber\\
&&\quad=
-\Spv\cdot\sov
+2\left(
\Spv_{13}\cdot\sov_{13}+
\Spv_{32}\cdot\sov_{32}+
\Spv_{21}\cdot\sov_{21}\right),
\label{EQ:correction}
\end{eqnarray}%
where $\Spv\equiv\Spv_{13}+\Spv_{32}+\Spv_{21}$, 
and   $\sov\equiv\sov_{13}+\sov_{32}+\sov_{21}$. 
   When $U$ in the SOI
is a superposition of spherically-symmetric ionic potentials, 
the vectors $\sov_{jk}$ have a transparent geometrical meaning, 
and are proportional to the triangle area $\arQ$.  In this case,  
$\sov_{jk}=a_{jk}\left(\pov_j-\pov_h\right)\times
                 \left(\pov_k-\pov_h\right)=a_{jk}{\bf\arQ}$.  
Then the SOI-generated shift in the carrier eigenenergies 
has the Dzyaloshinski-Moriya form~\cite{REF:Dzyaloshinski}.  By 
incorporating the shifts~(\ref{EQ:correction}), together with 
the Pancharatnam phase, we arrive at the elementary AH conductivity 
\begin{equation}
\sigma^{(1)}_{\rm AH}=  
\spv_{1}\cdot(\spv_{2}\times\spv_{3})
\sum\nolimits_{j}\delta\varepsilon_{j}\,
\partial G/\partial\varepsilon_{j}.
\label{EQ:AHEC}
\end{equation}%
As discussed above, Eq.~(\ref{EQ:AHEC}) 
has a nonzero macroscopic average, owing to the presence of a 
characteristic Pontryagin charge constructible from the $\Spv_{jk}$, 
that feature in the energy shifts, and the magnetization direction. 
A second consequence of the SOI-generated carrier-energy 
shift~(\ref{EQ:correction}) leads to the second contribution,  
$\sigma^{(2)}_{\rm AH}$.  Due to the feedback of the (fast) carrier 
freedoms, which provide an effective potential for the (slow) spin 
system, determined by Eq.~(\ref{EQ:correction}), the 
equilibrium probabilities of spin configurations having opposing 
Pancharatnam fluxes will no longer be equal.  (For this contribution, 
which is related not to $\partial G/\partial\varepsilon_{j}$ but to 
$G$ itself, there is no need to account for SOI-induced carrier-energy 
shifts in the current now being averaged over a nonsymmetric 
spin-configuration distribution.)\thinspace\  A contribution with
this origin has also been considered in Ref.~\cite{REF:Ye1999}.  
$\sigma^{(1)}_{\rm AH}$ and $\sigma^{(2)}_{\rm AH}$ are of the 
same order of magnitude. 

We now consider the question of how the physics of elementary triads 
relates to the macroscopic properties of manganites.  For hopping 
conductivity, the pathways taken by the current depend sensitively on 
the details of the core-spin configuration, and regions having certain 
local spin configurations will tend to be avoided by the current. 
This fact renders rather subtle the spin-configuration averaging 
procedure, which must also account for effects such as local spin 
correlations and excitations of various types.   
Let us try to identify which triads the AH current tends to favor. 
To favor their participation in the conducting network, the three 
core spins in the triad should at least have positive components  
along the magnetization direction.
For magnetic compatibility with its neighbors, the net magnetization 
of the triad should be roughly that of the bulk. 
Furthermore, to contribute appreciably to the AH current, the triad
should be as splayed as possible, given the above constraints.  This
favors symmetrical configurations of the triad spins; we call these
triads optimal triads.
As we shall see in the companion Letter~\cite{REF:Salamon}, these 
observations allow us to explain the striking experimental finding 
that the Hall resistivity depends on the magnetic field and 
temperature {\it only\/} through the magnetization and, moreover, 
to predict the explicit form of this dependence.

\noindent
{\sl Spin-orbit quantal phase and AHE in nonmagnetic triads\/}: 
We conclude with a remark concerning the hopping AHE in systems 
with nonmagnetic ions (in which case no Pancharatnam phases arise).  
In this case, the SOI quantal phase itself leads to an AHE.  
In the presence of the SU(2) gauge potential ${\bf A}_{\rm so}$
electrons moving around a nonmagnetic triad acquire a full SU(2) 
phase, not projected due to Hund's Rules.  Due to carrier-spin 
polarization, this phase leads to an AH current in the same way 
that the AB flux leads to the  OHE~\cite{REF:Holstein1}.  We 
emphasize that that the OH and AH effects in such systems should 
be experimentally distinguishable from one another.  For example, 
the AHE in the hopping regime should be observable in inversion 
layers of doped semiconductors in the absence of 
magnetic field, when electron spin-polarization 
is induced by circularly polarized light.   

\noindent
We thank I.~L.~Aleiner and V.~L.~Pokrovskii for helpful discussions, 
and authors of Ref.~\cite{REF:Ye1999} for communicating preliminary results 
of their work. This work was supported by DOE Grant DEFG02-96ER45439.
\vspace*{-0.5cm}	
  
\end{multicols}

\begin{references}
\bibitem{REF:numphon} 
By a two-phonon process (assisting the HE) we mean one involving 
probabilities (computed in perturbation theory) characterized by 
4 carrier-phonon interaction vertices; 
the probability of a single-phonon process (leading to the 
longitudinal conductivity) involves 2 such vertices.
\end{references}
\end{document}